\documentclass[aps,prb,twocolumn,groupedaddress]{revtex4}

\usepackage{gensymb}
\usepackage{multirow}
\usepackage[dvipdf]{graphicx}
\usepackage{rotating}
\usepackage{natbib}
\usepackage{xcolor}

\begin{document}

\title{Theoretical Prediction of Stable Cluster-assembled CdSe Bilayer and its Functionalization with Co and Cr Adatoms}

\author{Deepashri Saraf}
\author{Anjali Kshirsagar}
\email{anjali@physics.unipune.ac.in}
\affiliation{Department of Physics, Savitribai Phule Pune University, Pune 411007, India}
   

\begin{abstract} 

In this article, we present our results on bilayers assembled upon strategic
placement of Cd$_6$Se$_6$ clusters. These bilayers are studied for their
stability and electronic structure with the help of density functional theory
and are further analyzed using Bardeen, Tersoff and Hamann formalism for their
tunneling properties. Our calculations show that the hexagonal arrangement of
these clusters prevails as the most stable geometry showing all real phonon
modes. First-principles molecular dynamics studies on this 2D structure show
that it remains intact even at room temperature. This bilayer shows an indirect
semiconducting band gap of 1.28~eV with the current-voltage (I-V)
characteristics similar to a tunnel diode. Further, we functionalized this
bilayer using transition metal atoms, Co and Cr. The aim was to seek whether
the bilayer sustains magnetism and how the concentration affects its electronic
and magnetic properties. Co functionalization brings ferromagnetic ordering in
the bilayer which turns near half-metallic upon increasing the concentration.
On the other hand, Cr functionalization shows a transition from antiferro- to
ferromagnetic ordering upon increasing the concentration. The I-V
characteristics of all these functionalized bilayers show negative differential
conductance similar to a tunnel diode.

\end{abstract}

\maketitle

	\section{Introduction}

Cluster-assembled materials, in which atomic clusters are arranged methodically
in a periodic array to form a solid, are emerging materials for their
technological applications. Also known as nanostructured materials, they allow
the integration of multiple length scales into a hierarchical material
\cite{Khanna}. With precisely controlled building-blocks, such synergistic
cluster-assemblies can be used to create multitudes of periodic structures
having unusual symmetries that are ``custom-made'' for specific requirements.
Although various successful attempts have been made to explore the possibility
of existence of such materials \cite{Khanna}, smaller sized clusters are 
difficult to handle experimentally. Thus, modeling these materials
theoretically, generates an impetus for experimental processes.

Modern era of materials research revolves around 2-dimensional (2D) materials
as a consequence of discovery of graphene \cite{novo, allen}.  Recent times
have witnessed an emergence of variety of different 2D materials that have not
only opened new avenues of fundamental research but also of device designs.
Amidst the deluge of 2D materials, pseudo-planar and van der Waals (vdW) bonded
layered materials have maintained their own identity \cite{Molle, Jariwala_1}.
Keeping with the current research trends, we planned to explore the
cluster-assembled vdW bonded bilayers of CdSe consisting of buckled
pseudo-planar sheets. Also, having predicted the existence of such bilayer
theoretically, we further wanted to see whether the bilayer can form ordered
spin structures, upon introduction of transition metal (TM) atoms.  We
speculate that such materials can facilitate a number of potential
functionalities that can be readily engineered. Earlier such attempts were made
by Liu \textit{et al.} in their studies of cluster-assembled sheets of
endohedrally doped Si$_{12}$ clusters with vanadium atoms \cite{Liu_1}. They
observed that two different types of cluster-assembled sheets prefer
ferromagnetic ordering with free-electron-mediated mechanism.

This article presents our results on bilayers assembled using Cd$_6$Se$_6$
clusters that we further functionalize with the help of TM atoms, Co and Cr. We
also simulate scanning tunneling spectroscopy results by calculating the
tunneling properties of these pristine and functionalized bilayers with the
help of Bardeen, Tersoff and Hamann (BTH) formalism \cite{Bardeen, TerHam}
combined with first-principles density functional theory (DFT) approach.

	\section{Computational details}

Our calculations for Cd$_6$Se$_6$ cluster-assembled bilayers are based on DFT
formalism as implemented in Vienna Ab-initio Simulations Package (VASP)
\cite{VASP}. The structural and electronic properties are obtained with the
help of exchange-correlation energy functional as given by Perdew, Burke and
Ernzerhof (PBE) with projector-augmented-wave (PAW) method used to describe the
core electrons \cite{pbe,paw}. The self-consistent convergence criterion of
energy is set to 10$^{-5}$~eV. The occupation numbers are treated according to
the Gaussian scheme with a broadening of 0.001~eV. The bilayers are constructed
(in the xy-plane) in such a manner that the two periodic images are separated
by a distance of 15~{\AA} (in the z-direction). In order to optimize the
structures, relaxation procedures are carried out according to conjugate
gradient algorithm. The relaxation is achieved with the help of
Hellmann-Feynman forces and the stress tensors by taking appropriate
derivatives of total energy of the unit cell in every iteration with the
convergence threshold on forces set as 0.01~eV/{\AA}. The Brillouin zone (BZ)
is represented by the Monkhorst-Pack \textbf{k}-point mesh of
$8\times8\times1$. For weak interactions involved in the computation, due to
the presence of bilayered structures, we have used PBE+D3 method with Grimme
vdW corrections as implemented in VASP \cite{VDW}. To understand the structural
stability, phonon bandstructure and density of states (DOS) are calculated
using density functional perturbation theory with the help of linear response
method \cite{phonopy}.

In order to study the tunneling properties of these pristine and functionalized
bilayers, we use BTH formalism as implemented by He \textit {et al.}
\cite{pandey} in their scanning tunneling microsope (STM)-like setup. As per
this formalism, the electron tunneling current can be calculated in the
low-bias limit using first-order perturbation theory as follows:

\begin{eqnarray}
\hskip -2cm
I &=& \frac{4\pi e}{\hbar} \int_{-\infty}^{+\infty}\rho_s \left( \epsilon + \frac{eV} {2}\right) 
\rho_t \left( \epsilon - \frac{eV} {2}\right) \nonumber \\ 
&& \times e^{-2d \sqrt {2(m/\hbar^2)(\phi_{av} - \epsilon)}} \nonumber \\
&& \times \left\{ \left[ f \left( \epsilon - \frac{eV}{2} \right) \right]
\left[ 1 - \left[ f \left( \epsilon + \frac{eV}{2} \right) \right] \right] \nonumber \right. \\ 
&& \:\:\: \left.-\left[ f \left( \epsilon + \frac{eV}{2} \right) \right] \left[ 1 + \left[ f
\left( \epsilon - \frac{eV}{2} \right) \right] \right] \right\} d\epsilon \nonumber
\end{eqnarray}

\noindent where $\rho_s$ and $\rho_t$ are the projected densities of states
(PDOS) of the sample and the tip respectively, $d$ is the tip to sample
distance, $\epsilon$ is the injection energy of the tunneling electron, $e$ is
the electronic charge, $m$ is the effective mass of the electron, $\hbar$ is
the Planck constant, $\phi_{av}$ is the average work-function of the sample and
the tip and $f$ is the Fermi distribution function. In this particular
formalism, due to the low-bias criterion $m$ and $\phi_{av}$ are assumed to be
constant. Since, the tip and the sample are assumed to be in electrochemical
equilibrium, their Fermi energies are aligned and are taken to be the reference
energy in the above equation. Bias-induced changes are not included on the
sample DOS, which occur only at high applied bias. In our STM-like setup, the
sample is the bilayer (pristine/functionalized), while the tip is modeled by
7-atom gold cluster. This fully relaxed tip geometry is chosen to mimic the
sharp STM tip. The DOS of the STM-tip is artificially broadened (broadening
factor 0.2~eV) to consider the broadening due to the semi-infinite nature of
the tip. While designating the value of broadening, the reported value for life
time broadening of electrons in a cluster in a scanning tunneling spectroscopy
study is taken into account \cite{BUK}.

	\section{Results and discussion}

	\subsection{Geometry and electronic structure of CdSe bilayers}

\begin{figure}[ht]
\begin{center}
        \includegraphics[width=0.3\textwidth]{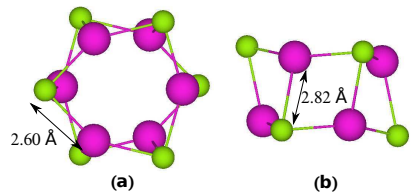} 
\end{center}
\caption{Geometry of Cd$_6$Se$_6$ cluster. (a)~Top view and (b)~Side view. Cd 
and Se atoms are indicated in magenta and green respectively. The colour scheme is 
maintained throughout this article.
\label{fig:1}}
\end{figure}

Motivation behind this study is to examine whether an assembly of small CdSe
clusters can form stable 2D-sheets that can further be used in different
applications. As a first step towards this, we chose Cd$_6$Se$_6$ cluster which
is the smallest wurtzite cage with $C_1$ symmetry as observed by Jose
\textit{et al.} \cite{jose}. The cluster consists of two planar Cd$_3$Se$_3$
clusters stacked on top of each other in chair conformation, where each
Cd$_3$Se$_3$ cluster forms a hexagon of side 2.60~{\AA} and the Cd-Se stacking
distance is 2.82~{\AA} (see Figure \ref{fig:1}).

\begin{figure*}[ht]
\begin{center}
        \includegraphics[width=0.7\textwidth]{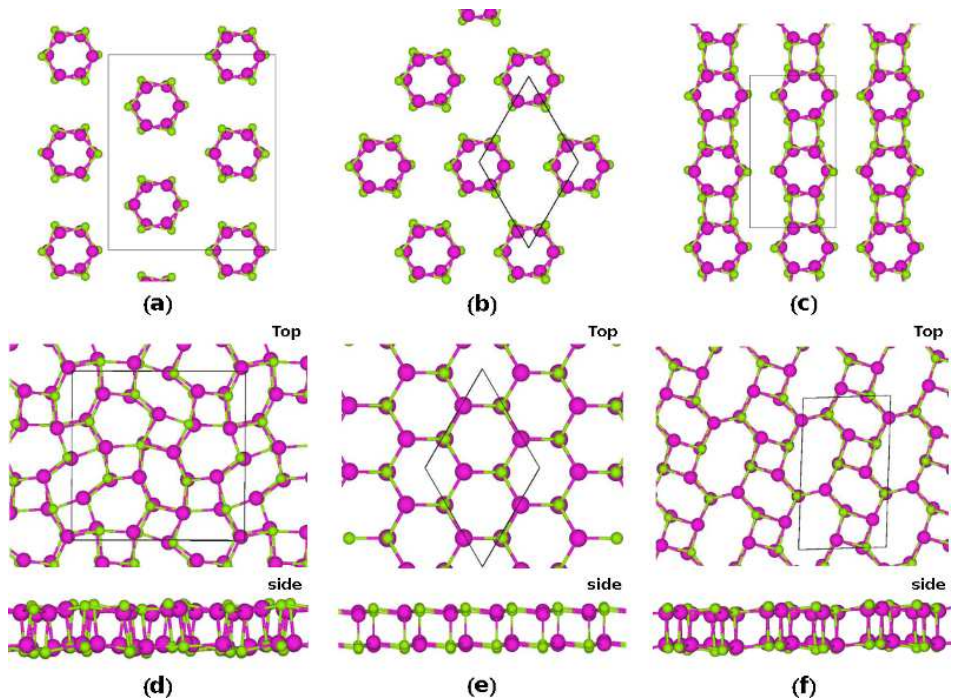}
\end{center}
\caption{Initial and optimized geometries of Cd$_6$Se$_6$ cluster-assembly, for
Type-1 (a) and (d), Type-2 (b) and (e) and Type-3 (c) and (f) structures, respectively.
\label{fig:2}}
\end{figure*}

There may be a large number of possible 2D configurations to be found by
arranging Cd$_6$Se$_6$ clusters as building blocks. Here we design three
configurations and let them undergo unconstrained relaxation. These
configurations are shown in the upper panel of Figure \ref{fig:2}~(a), (b) and
(c), and are named as Type-1, Type-2 and Type-3 structures. The corresponding
geometries of these configurations upon relaxation are shown in Figure
\ref{fig:2}~(d), (e) and (f). In each of these cases, the clusters form
bilayered sheets that show buckling in both the layers. Out of these three
structures, Type-2 has the smallest number of atoms in the unit cell (6 Cd and
6 Se atoms) with a hexagonal lattice (7.83~{\AA}) i.e. the unit cell consists
of a single Cd$_6$Se$_6$ unit. Unit cell of Type-1 bilayer consists of four
units of Cd$_6$Se$_6$ cluster (lattice parameters, $a = 14.09~{\AA}$ and $b =
14.39~{\AA}$) and that of Type-3 bilayer consists of two units of Cd$_6$Se$_6$
cluster (lattice parameters, $a = 7.94~{\AA}$ and $b = 13.80~{\AA}$). Average
width of the bilayer in Type-1 (2.78~{\AA}) mostly remains uniform with the
most uneven distribution of in-plane Cd-Se bond lengths ranging from 2.65~{\AA}
to 2.78~{\AA}, amongst the three bilayers. For Type-2 structure, the bilayer
width varies between 2.82~{\AA} and 2.91~{\AA}. The average Cd-Se bond length
in this case is 2.64~{\AA} and the hexagons formed by Cd-Se atoms are not
regular. Average width of the bilayer in Type-3 geometry is 2.82~{\AA} with
in-plane Cd-Se bond lengths varying from 2.62~{\AA} to 2.78~{\AA}. Comparing
all the relaxed geometries we observe that Type-2 structure has the highest
symmetry amongst the three types and it shows similar structure to that of
$AA'$ type stacking in $h$-BN bilayers \cite{heine}. $AA'$ is shown to be the
most stable stacking order in $h$-BN bulk and bilayer forms, both
experimentally as well as theoretically. Type-1 structure shows a bilayer
composed of hexagons and quadrilaterals, whereas optimized geometry of Type-3
structure contains a stretched array formed of octagons and hexagons.

\begin{figure}[ht]
\begin{center}
        \includegraphics[width=0.45\textwidth]{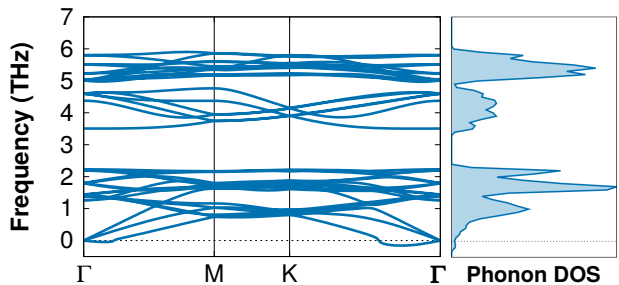}
\end{center}
\caption{Phonon bandstructure and DOS of Cd$_6$Se$_6$ cluster-assembly, for Type-2 structure.
\label{fig:3}}
\end{figure}

The relaxed geometries were further compared energetically to understand their
relative stabilities. We calculated the binding energy ($B.E.$) per atom as
follows,

\begin{equation}
B.E./atom = \frac {E_{Total} - \{n \times (E_{Cd} + E_{Se})\}} {n}
\end{equation} 

\begin{table}
\begin{center}
\begin{tabular}{c c}
\hline
~\\
Structure       &       $B.E./atom$     \\
		&	(eV)		\\
\hline \hline
~\\
Type-1          &       -2.67           \\
~\\
Type-2          &       -2.79           \\
~\\
Type-3          &       -2.69           \\
~\\
\hline
\end{tabular}
\end{center}
\caption{Comparison between the binding energies of Type-1, Type-2 and Type-3 structures.
\label{table:1}}
\end{table}

The values of $B.E./atom$ are shown in Table \ref{table:1} for all the three
structures and they differ only marginally. It can be seen that Type-2
structure, which has the highest symmetry also has the lowest $B.E.$ value.
This indeed reiterates that layered materials are predominantly formed in
hexagonal symmetries, including different stacking orders of the hexagonal
layers \cite{heine}. To confirm the stability of Type-2 assembly (owing to its
lowest $B.E.$ value), we calculated its phonon bandstructures that is shown in
Figure \ref{fig:3}. With the exception of small imaginary frequencies around
the $\Gamma$ point of the BZ the phonon bandstructure does not show any
unstable phonon modes confirming its stability. Small imaginary frequencies
(\textless 10~THz) can be attributed to the limitations of the numerics. \\

Further, it is imperative to study whether the predicted 2D
material withstands small temperature fluctuations. Hence, we performed
$ab~initio$ molecular dynamics (MD) simulations at room temperature i.e. 300K
for $\sim$ 10000 fs. We observed that apart from some small thermal
fluctuations the Type-2 structure remains in tact. Thus, our observed lowest
energy cluster-assembled 2D bilayer of CdSe, is thermally stable upto 300K. \\

\begin{figure*}[ht]
\begin{center}
        \includegraphics[width=0.85\textwidth]{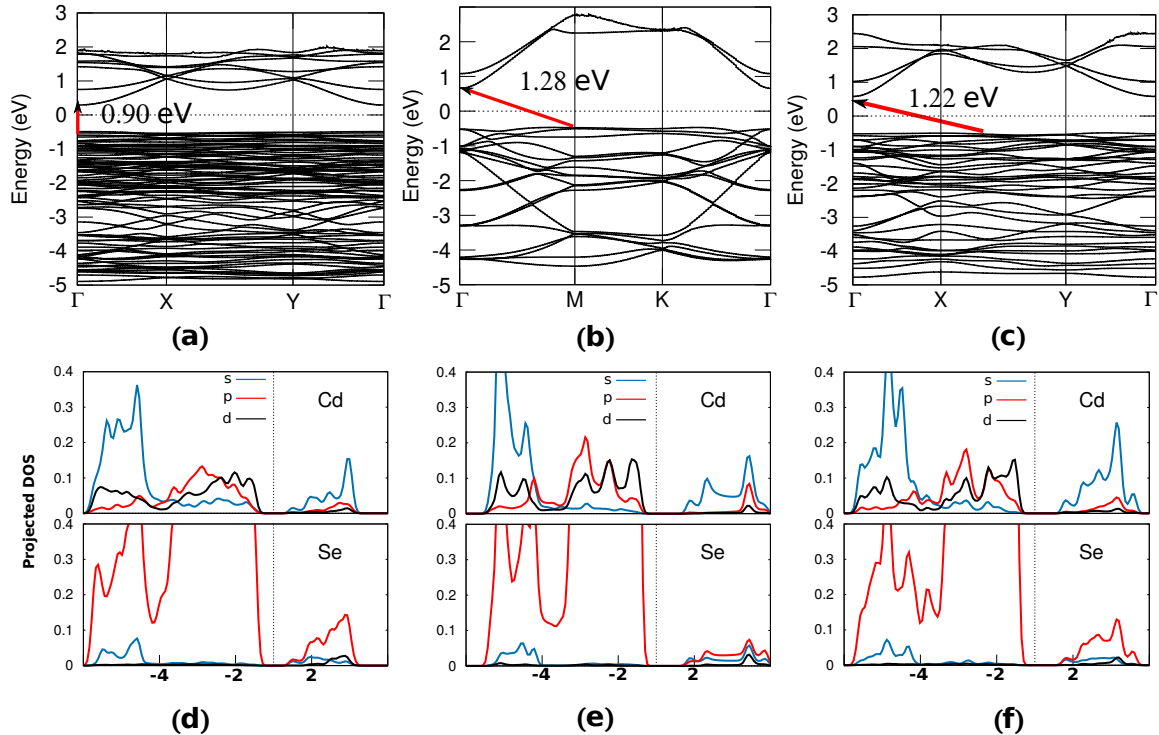}
\end{center}
\caption{Electronic structure and PDOS of Cd$_6$Se$_6$ cluster-assembly are shown for
Type-1 structure in (a) and (d), for Type-2 structure in (b) and
(e) and for Type-3 structure in (c) and (f), respectively.
\label{fig:4}}
\end{figure*}

Figure \ref{fig:4} shows the electronic bandstructures and PDOS of Cd$_6$Se$_6$
cluster-assembled bilayers. Type-1 structure has a direct band gap of 0.90~eV
at $\Gamma$ point of the BZ, whereas Type-2 structures has an indirect band gap
of 1.28~eV from $M \rightarrow \Gamma$. Type-3 structure also shows an indirect
band gap of 1.23~eV where the conduction band minimum (CBM) lies between $X$
and $Y$ points and the valence band maximum (VBM) lies at the $\Gamma$ point of
the BZ. In all the three structures, the VBM consists of Se $p$ and Cd $d$
states with Se $p$ being predominant and the CBM is made up of Cd $s$ states.
CdSe, in its bulk form has been reported to have experimental band gap of
1.84~eV at 0~K \cite{kittel}, whereas our calculations for bulk wurtzite form
of CdSe show a band gap of 0.53~eV. This happens due to the tendency of DFT to
underestimate the band gaps and we expect that just as the case of bulk CdSe,
the actual band gaps for our bilayers, would even be higher than our reported 
values. Hence, the cluster-assembled bilayers, studied in the current work, may
show band gaps larger than that of bulk CdSe.

	\subsection{Transition metal doped CdSe bilayers}

We now present the effects of TM adatoms in our Type-2 CdSe bilayer that we
obtained as the most stable cluster-assembled sheet. TM doping in a
semiconductor is a viable way to improve its properties as a probable material
for spintronic device. Group II-VI semiconductors doped with different TM atoms
have been investigated as diluted magnetic semiconductors by various
researchers. Co doping is known to induce antiferromagnetic behaviour in bulk
CdSe \cite{Niu} whereas Cr doping in II-VI semiconductors shows ferromagnetic
behaviour \cite{Niwa, Shri}. For our study, we investigated their effects on
structural, magnetic and electronic properties of the bilayer. There are two
fundamentally different locations for the introduction of adatoms: one
in-between the layers and other on the top. Our calculations indicate that upon
insertion of TM adatom in between the layers, our structure gets highly
distorted resulting in non-Cd$_6$Se$_6$ configuration. Therefore we report only
the later case hereafter. We present two cases of Co/Cr adsorbed on top of
Type-2 Cd$_6$Se$_6$ cluster-assembled bilayer. In the first one, single TM atom
is kept on top of center of every hexagon giving 50\% adsorption per Cd-Se pair
in the bilayer, while in the second case a single TM atom is kept on top of
center of alternate hexagon giving $\sim$17\% adsorption per bilayer atom. With
these initial geometries we performed unconstrained relaxation to obtain the
minimum energy solution. Additionally, magnetic ground state is determined by
unconstrained minimization of all possible magnetic configurations. Below we
report only the ground state magnetic configuration of 50\% and 17\%
concentration of Co and Cr adatoms over Type-2 CdSe bilayer.

	\subsubsection{Structural, electronic and magnetic properties : Co-adatom}

\begin{figure*}[ht]
\begin{center}
        \includegraphics[width=0.6\textwidth]{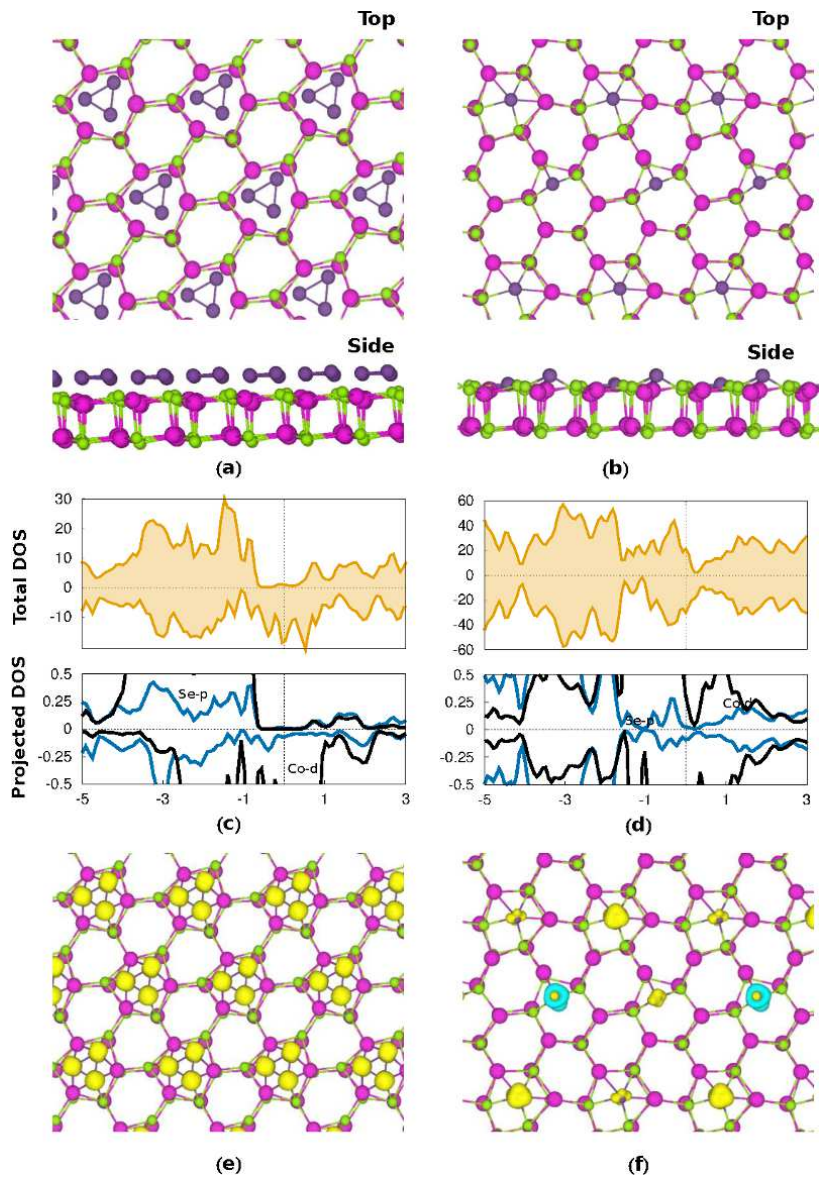}
\end{center}
\caption{Side and top views of the optimized geometries of Type-2 Cd$_6$Se$_6$
cluster-assembled bilayer with Co adsorption at (a)~50\% and (b)~17\%
adsorption per Cd-Se pair in the bilayer. Co atoms are indicated in purple and
the colour is maintained throughout this article. Total DOS and PDOS of the
same bilayers with Co adsorption are shown in (c)~at 50\% and in (d)~at 17\%
per Cd-Se pair in the bilayer. Se $p$ and Co $d$ states are indicated in blue
and black respectively. Spin densities of the bilayers with Co adsorption are
shown in (e) and (f) at 50\% and 17\% adsorption per Cd-Se pair in the bilayer
respectively. The isosurfaces shown in the figure are taken at one fourth of
the maximum isovalue. The up-spin and the down-spin densities are shown in blue
and yellow respectively.
\label{fig:5}}
\end{figure*}

We notice that for higher concentration of Co-adatom (50\%), a tendency of
clustering of Co atoms is seen over the bilayer (see Figure \ref{fig:5} (a)).
Crossing over the underlying bonds Co atoms form triangular geometries over
every alternate hexagon along the armchair direction. The distorted triangle
has a mean bond length of 2.23~{\AA}, and is located at a distance of
$\sim$2.11~{\AA} above the bilayer. The underlying bilayer also distorts to
accommodate the adatoms. The resulting system is ferromagnetic with a magnetic
moment of 10.19 $\mu_B$/unit cell and has a $B.E.$ per atom of -3.73~eV. This
behaviour is also seen from the total DOS and PDOS in Figure \ref{fig:5}~(c).
Spin-up DOS for this configuration shows a gap near Fermi energy, whereas the
spin-down DOS shows conducting nature resulting in half-metalic behaviour.
Spin-down channel mainly conducts through Co $d$ states, that are empty in
spin-up channel. Se $p$ states also acquire small magnetization due to the
presence of Co atoms. This bahaviour is different than bulk where Co impurities
are antiferromagnetically coupled. Antiferromagnetic state of the same
configuration lies 0.58~eV above the ferromagnetic state.

\begin{figure*}[ht]
\begin{center}
        \includegraphics[width=0.6\textwidth]{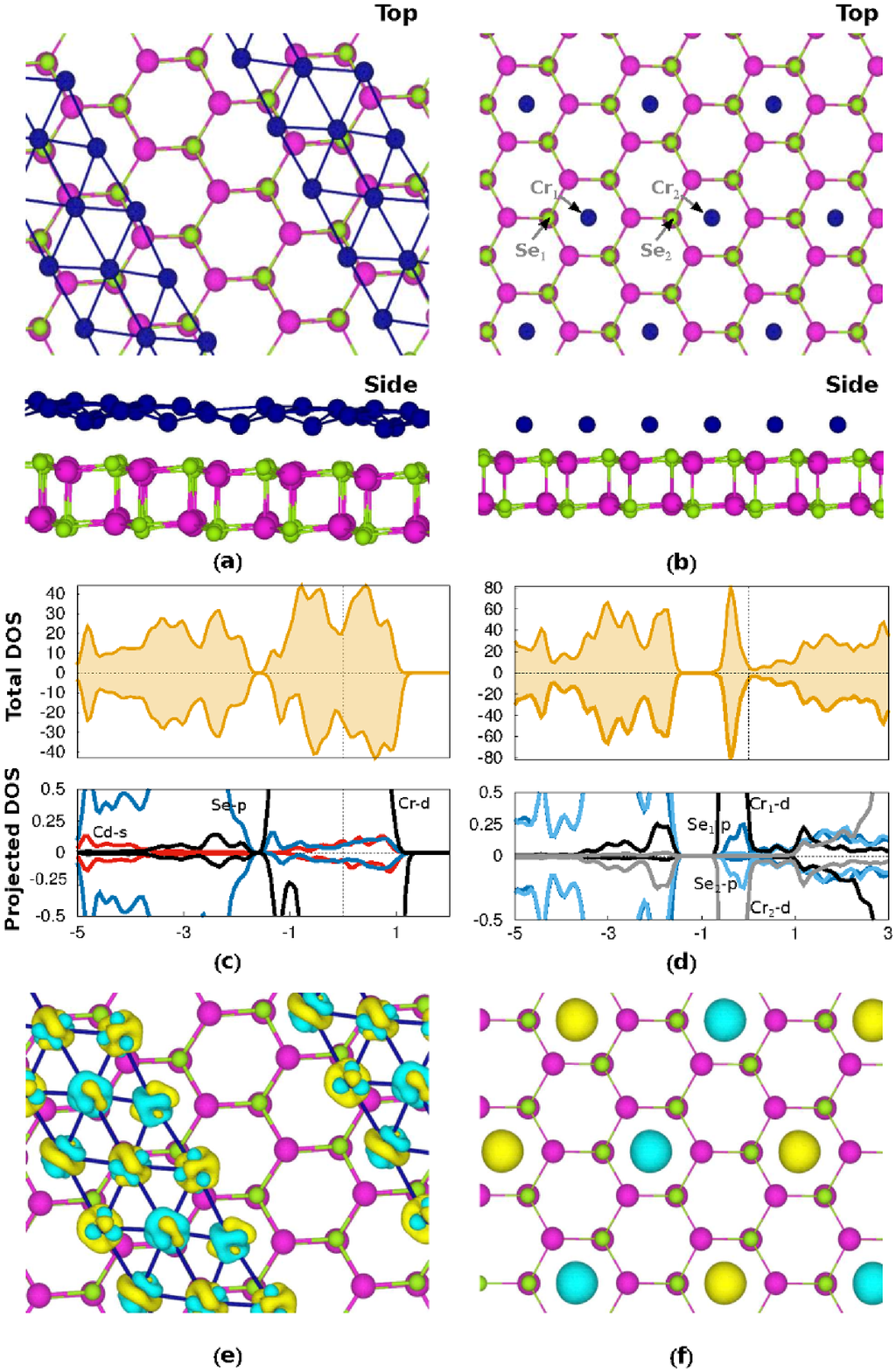}
\end{center}
\caption{Side and top views of the optimized geometries of Type-2 Cd$_6$Se$_6$
cluster-assembled bilayer with Cr adsorption at (a)~50\% and (b)~17\%
adsorption per Cd-Se pair in the bilayer. Total DOS and PDOS of the same
bilayers with Cr adsorption at (c)~50\% and (d)~17\% per Cd-Se pair in the
bilayer. Cd $s$, Se $p$ and Co $d$ states are indicated in red, blue and black
respectively. In Figure (d), we have additionally plotted the PDOS for two
different sites of Se and Cr atoms in light blue and grey respectively, that
clearly demonstrates the antiferromagnetic coupling. The sites chosen are
indicated in Figure (b). Spin densities of the bilayers with Cr adsorption at
(e)~50\% and (f)~17\% adsorption per Cd-Se pair in the bilayer. The isosurfaces
shown in the figure are taken at one fourth of the maximum isovalue. The
up-spin and the down-spin densities are shown in blue and yellow respectively.
\label{fig:7}}
\end{figure*}

Upon reducing the concentration of Co to 17\%, as seen from Figure
\ref{fig:5}~(b), similar to the previous case, the adatoms move away from their
initial positions and the underlying bilayer distorts. Similar to its 50\%
counterpart, the TM atoms are ferromagnetically coupled and the system has a
$B.E.$ of -2.87~eV/atom. The magnetic moment now reduces to 1.00 $\mu_B$/unit
cell. Co atoms have an average distance of 7.42~{\AA} between each other. The
ferromagnetic nature is also seen from the total and site projected DOS plots
shown in Figure \ref{fig:5}~(d). We can see the presence of gap states near
Fermi energy in both spin-up and spin-down channels that originate mainly due
to the presence of Co $d$ states. Thus, one can see the possibility to tune the
band gap and magnetism using concentration of Co. Moreover one can also
speculate that certain concentration ($\sim$50\%) of Co-adatom may give rise to
half-metallic phases - a feature that has been rarely investigated in 2D
materials.

	\subsubsection{Structural, electronic and magnetic properties : Cr-adatom}

\begin{figure*}[ht]
\begin{center}
        \includegraphics[width=0.6\textwidth]{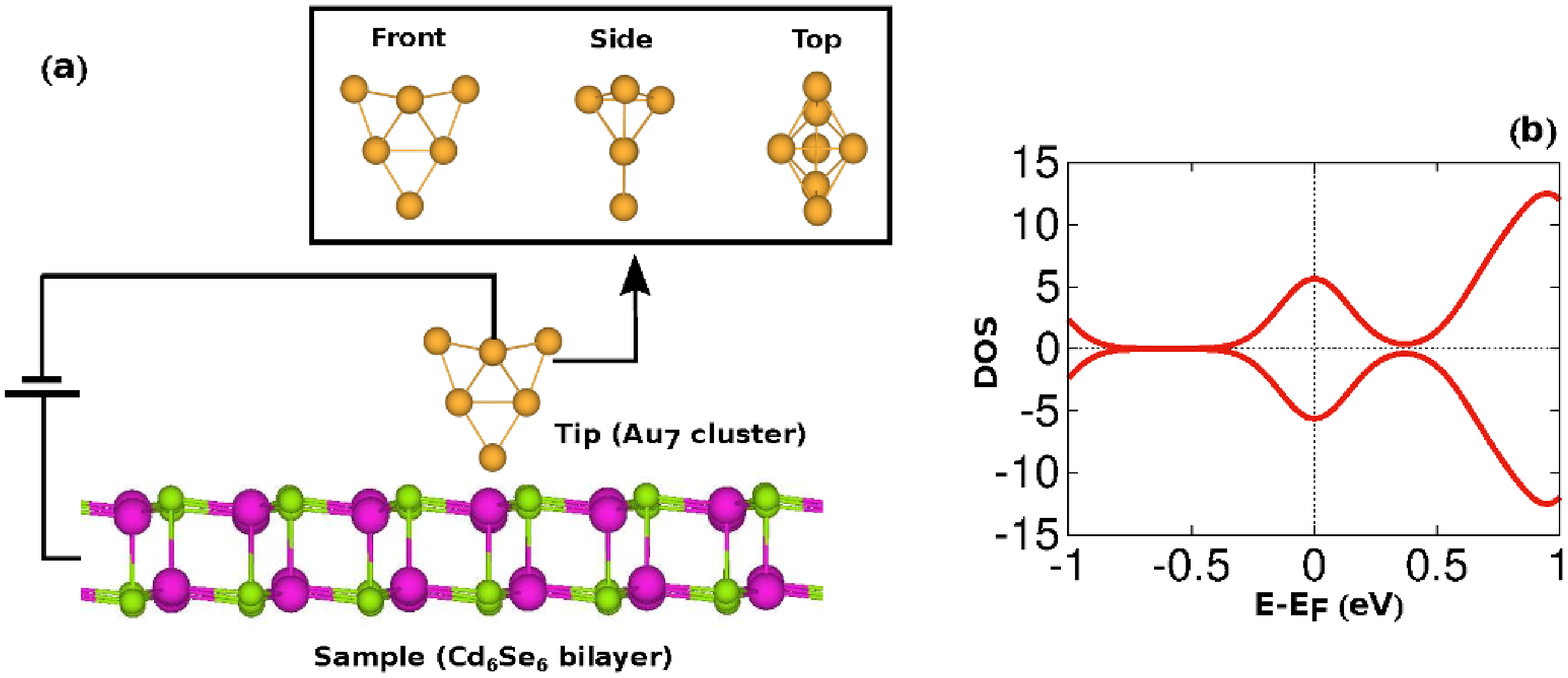}
\end{center}
\caption{(a)~Schematic illustration of the STM-like set-up having Type-2
Cd$_6$Se$_6$ bilayer as the sample with Au probe tip. Inset shows the geometry
of the Au probe tip from various angles. (b)~DOS plot of the Au probe
tip.
\label{fig:10a}}
\end{figure*}

Cr-adsorption at 50\% concentration, upon full relaxation, also shows
clustering tendency, but unlike Co where atoms form islands, here atoms form
ribbons. As seen in Figure \ref{fig:7}~(a), Cr adatoms form ribbons of width
5.07~{\AA} along the armchair direction resulting in an array of triangular
structures. The mean bond length in the triangle is 3.60~{\AA} and two stripes
are separated by 9.45~{\AA}. Unlike Co-adsorption, small degree of distortion
is observed in the underlying CdSe bilayer. Triangular array obviously opens up
the possibility of several magnetic structures. We performed series of
calculations of ferromagnetic and antiferromagnetic ordering and found that the
ferromagnetic and antiferromagnetic states are barely separated in energy by
only 1~meV/atom with the value of $B.E./atom$ as -3.41~eV. The magnetic moment
in case of ferromagnetic phase is 1.00$\mu_B$/unit cell. In Figure \ref{fig:7}
(c), we present the total DOS and PDOS of the ferromagentic phase, which mainly
originates from Cr $d$ states with Cd $s$ and Se $p$ states showing their
presence by acquiring small magnetization. \\

On the other hand, lower concentration of Cr (17\%) prefers to be in
antiferromagnetic phase. As seen from Figure \ref{fig:7}~(b), the geometry of
the bilayer does not change much and despite being $\sim$7.84~{\AA} apart, Cr
atoms are coupled antiferromagnetically (See Figure \ref{fig:7} (d)). In the
figures of total DOS and PDOS (See Figure \ref{fig:7} (d)), the antiferromagnetic
coupling is indicated by showing $d$ states of two Cr atoms with black and
grey; and $p$ states of two Se atoms with blue and light blue. As discussed
before, for both the concentrations of Cr, the effective magnetic moment is
fairly small. This is also evident from the spin density plots of these
structures shown in Figure \ref{fig:7} (e) and (f). It is worth noticing that for
higher concentration, the magnetic order is destroyed. This behaviour is
opposite to the behaviour of bulk Cr-doped CdSe system. Thus we expect that
with further increase in concentration, system will become non-magnetic. Thus
of Cr offers a viable way to achieve desired magnetic phases. \\ 

The most noticeable difference between the two adatoms is that changing the
concentration of Co adatoms can tune the band gap and can help get desirable
magnetic moment whereas changes in the Cr concentration changes the magnetic
nature of the bilayer. Co adatom has strong binding which also results in
clustering of the adatoms and structural deformation of bilayer. Cr adatoms, on
the other hand forms highly ordered patterns, maintaining the structure of the
underlying bilayer.

	\subsubsection{Tunneling properties}

We now investigate the pristine and TM doped bilayers with the help of BTH
formalism. BTH formalism is well-suited in the limits of small bias voltage
(V), $e*V << \phi_m$, where $\phi_m$ is the work function of the tip.
Considering the typical metal work function, $\phi_m$ = 4~eV, bias voltage
range below 2~V can typically be useful in BTH formalism \cite{FunPico}. In our
current work, the STM-like setup consists of pristine/functionalized Type-2
Cd$_6$Se$_6$ bilayer as sample and 7-atom gold cluster as tip (see Figure
\ref{fig:10a}~(a)). This fully relaxed tip geometry is chosen to mimic the
sharp STM tip. The DOS of the STM-tip is artificially broadened (broadening
factor 0.2~eV) to consider the broadening due to the semi-infinite nature of
the tip. It should be noted that the Au$_7$ cluster that we are incorporating
as a tip has paramagnetic ground state that is also evident in the DOS of the
tip in Figure \ref{fig:10a}~(b).

\begin{figure*}[ht]
\begin{center}
        \includegraphics[width=0.7\textwidth]{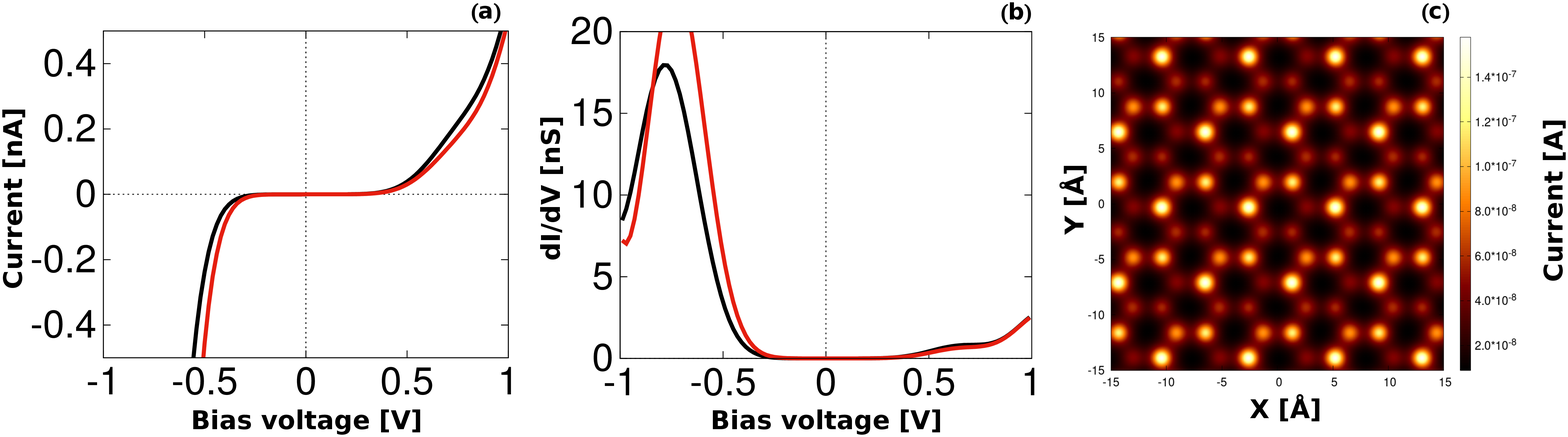}
\end{center}
\caption{Tunneling properties of Cd$_6$Se$_6$ cluster-assembled bilayer showing
(a)~I-V and (b)~dI/dV characteristics and (c)~constant height mode image.
Depending upon the position of the tip over the bilayer, the graphs are plotted
in red where the tip is held over a Cd atom and in black where the tip is held
over an Se atom. Bright regions in Figure (c) indicate Cd and Se atoms whereas
the dark regions indicate the voids.
\label{fig:11_a}}
\end{figure*}

In Figure \ref{fig:11_a} we present the calculated tunneling characteristics of
Type-2 Cd$_6$Se$_6$ bilayer for a low-bias range of -1.0~V to 1.0~V. The I-V
characteristics are computed at Cd and Se sites on the bilayer. By positioning
the tip exactly above these sites, we obtain the spatially resolved tunneling
spectra. These spectra are directly dependent on the sum of the relevant site
projected DOS of atoms in that plane and the adjoining planes till the
exponential distance factor in the tunneling current equation becomes
negligible. I-V characteristics for the bilayer are shown in Figure
\ref{fig:11_a}~(a). The curves are obtained upon sweeping the tip bias and
calculating the current at a particular lateral tip position, keeping the tip
to sample distance fixed ($\sim$4.5{\AA}). The I-V characteristics show
rectification nature since no states are available in the gap region (1.28~eV)
to contribute to the tunneling current. Figure \ref{fig:11_a}~(b) shows the
differential conductance (dI/dV) characteristics which correlates well with the
LDOS of the structure calculated using DFT. Changing the tip-sample separation
affects the magnitude of tunneling current, but the nature of graphs remains
unaffected. We also simulate the constant height mode STM images for the
bilayer as shown in Figure \ref{fig:11_a}~(c). These images provide the
topography of this quasi-2D structure. The height above the bilayer at which
the constant height mode image is taken, is chosen such that the optimum
resolution of the bilayer is obtained. The geometric features of the bilayer
are clearly seen in this image indicating the hexagonal patterns in the form of
bright spots (i.e. Cd and Se atoms). The brightness of the atoms shows their
relative proximity to the tip.

\begin{figure*}[ht]
\begin{center}
        \includegraphics[height=0.8\textheight]{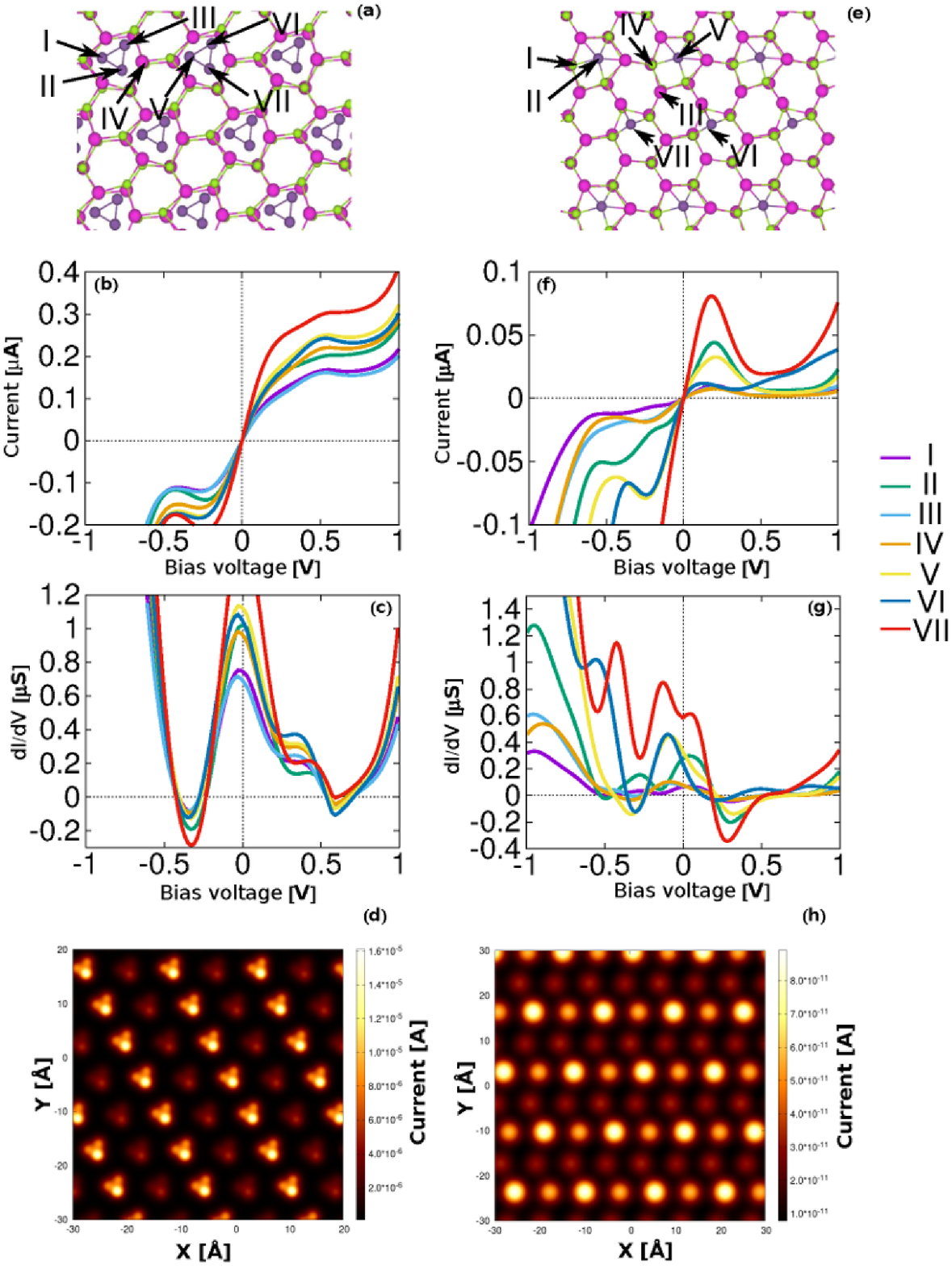}
\end{center}
\caption{Geometries of Cd$_6$Se$_6$ cluster-assembled bilayer with (a)~50\% and
(b)~17\% Co adsorption showing the numbering of the positions at which the STM 
tip is placed to calculate I-V ((b),(f)) and dI/dV ((c),(g)) characteristics
along with constant height mode images ((d),(h)) of the corresponding
structures. Bright regions in Figures (d) and (h) indicate atoms.
\label{fig:11}}
\end{figure*}

\begin{figure*}[ht]
\begin{center}
        \includegraphics[height=0.8\textheight]{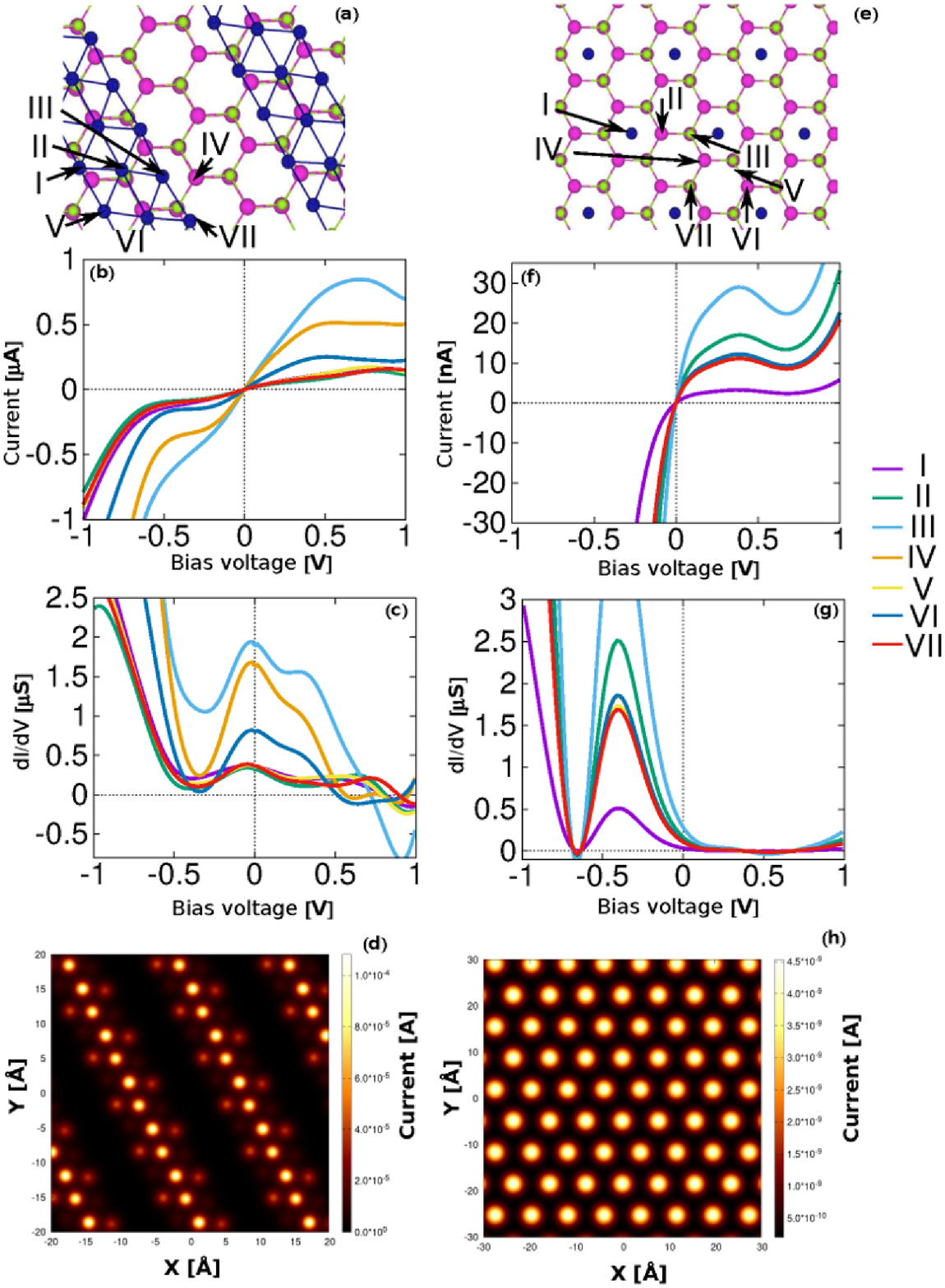}
\end{center}
\caption{Geometries of Cd$_6$Se$_6$ cluster-assembled bilayer with (a)~50\% and
(b)~17\% Cr adsorption showing the numbering of the positions at which the STM 
tip is placed to calculate I-V ((b),(f)) and dI/dV ((c),(g)) characteristics
along with constant height mode images ((d),(h)) of the corresponding
structures. Bright regions in Figures (d) and (h) indicate atoms.
\label{fig:12}}
\end{figure*}

We now focus our attention on the tunneling properties of the TM 
doped bilayer for different compositions. The results for these are compiled in
Figures \ref{fig:11} and \ref{fig:12} for Co doped and Cr doped bilayers
respectively.  In these figures, the tip positions are numbered and colour
coded. Graph for each position is then drawn with the same colour in all the
further figures depicting I-V characteristics corresponding to various
positions numbered as I to VII in Figures (b) and (f), dI/dV curves in Figures
(c) and (g) and constant height mode images in (d) and (h).

Let us first discuss the results for Co adsorbed bilayer presented in Figure
\ref{fig:11}. The I-V curves for all the positions for higher (Figure
\ref{fig:11}~(b)) as well as lower concentration (Figure \ref{fig:11}~(f))
maintain the linearity near zero bias, which is consistent with their
overlapping valence and conduction states seen in the DOS. It is known that the
tunneling current is directly related to the convolution of relevant regions of
the local density of states (LDOS) of tip and sample. Since there is finite
LDOS at Fermi energy we see increase in tunneling current with increase in
applied bias voltage. This behaviour is consistently present at every position
of 50\% Co-doped bilayer, but as the bias voltage reaches 0.5 V, we observe
that the tunneling current starts decreasing with an increase in the bias
voltage. The tunneling current decays upto a certain bias voltage range and
begins to rise again as bias increases further. This feature is known as
negative differential conductance (NDC) and it is characteristic of tunnel
diode. NDC is seen in the dI/dV characteristics which correlates well with the
LDOS calculated using DFT (see Figure \ref{fig:11}~(c)). In case of 17\%
Co-doped bilayer, the NDC arises at tip bias of $\sim$0.20 V and it is much
more pronounced than that in case of 50\% Co-adsorption. dI/dV characteristics
of 17\% Co-doped bilayer are seen in Figure \ref{fig:11}~(g). For both the
concentrations, the NDC features are also seen in negative bias region. NDC
indicates that no states are available for tunneling in the energy range
resulting in a little overlap of LDOS of Au-tip and the Co-doped bilayer,
therefore there is decrease in the tunneling current even with an increase in
bias voltage. The onset of NDC results from a sharp drop in the tunneling
probability when bias voltage reaches an energy level at which tunneling is
forbidden. For higher composition of Co-dopant (50\%), the higher density of TM
states allows the current to increase more-or-less in linear fashion. Reduction
in the Co-dopant concentration reduces the number of energy levels allowed for
transition, hence reducing the tunneling current and giving rise to more
pronounced NDC features. Constant height mode STM images are also depicted in
Figure \ref{fig:11}~(d) and (h). These images reveal the geometric features of
the Co-doped bilayers. In Figure \ref{fig:11}~(d), one can notice the
triangular patterns formed by Co atoms on top of the bilayer. But only the
atoms in the alternate rows of triangles are resolved well due to the fact that
the alternate triangles are placed slightly (~0.2 {\AA}) closer to the bilayer
than others. In Figure \ref{fig:11}~(h), Cd and Se atoms in the same plane as
that of Co are also observed as bright spots. 

Figure \ref{fig:12} shows the tunneling characteristics of the bilayer with Cr
adsorption. Similar to the Co-adsorption, one can observe the NDC features in
Cr-doped bilayer for both the concentrations. The NDC exists in the region of
bias voltage where there is minimum in the tip DOS. This is also confirmed in
dI/dV plots for both the concentrations (see Figures \ref{fig:12}~(c) and (g)).
Constant height mode STM images for Cr-doped bilayers are shown in Figure
\ref{fig:12}~(d) and (h). In Figure \ref{fig:12}~(d), the ribbon-like patterns
made by Cr atoms are clearly seen. The atoms that look diffused in these
ribbons are the ones that are not in the same plane. Similarly, in Figure
\ref{fig:12}~(h), one can see the Cr-atoms in the form of hexagonal patterns. 

	\section{Conclusions}

To summarize, we predicted and analyzed novel quasi-2D structures of CdSe built
intuitively using clusters as building blocks. We also functionalized them to
demonstrate their ability to control their electronic and magnetic properties.
The most stable configuration of CdSe bilayer shows an indirect band gap of
1.28~eV and I-V characteristics of a schottky diode. The bilayer
shows dynamical stability via all real phonon modes and sustains the
temperatures upto 300K as evident from our $ab~initio$ MD calculations.

We used TM atoms, Co and Cr, to functionalize the most stable configuration of
the bilayer. We found that even small concentration of Co adatoms, reorganizes
the bilayer to make it ferromagnetic. Upon increasing the Co concentration,
system shows tell-tale signs of half-metallic systems. It remains to be seen if
pure half-metallic behaviour can be obtained for certain concentration. On the
other hand, Cr doping shows a transition from ferro to antiferromagnetic
ordering upon decreasing the adatom concentration. The exciting effect of
doping is seen in the I-V characteristics, which we compute using BTH
formalism. While pristine bilayer shows classic Schottky diode-like features,
we found that the functionalized system shows the characteristics of a tunnel
diode via NDC.

We expect that our studies bring out the versatile nature of the
cluster-assembled CdSe bilayer that can find novel applications in the field of
electronics and spintronics.

AK acknowledges the financial support from the Nanomission Council, Department
of Science \& Technology, Government of India (Grant code: SR/NM/NS-15/2011(G))
through a major research project and DST-PURSE and DST-FIST grants to
Savitribai Phule Pune University and Department of Physics, respectively. DS
acknowledges financial support from Universities Grant Commission - Basic
Scientific Research (UGC-BSR) through a fellowship. We also thank C-DAC, Pune
for use of their computing facilities. The geometries of all the structures
shown in this article are generated using VESTA \cite{vesta}. 

\bibliography{biblio}
\end{document}